\documentclass[
superscriptaddress,
 amsmath,amssymb,
 aps, 
pra,
]{revtex4-2}
\usepackage{amsmath}
\usepackage{amssymb}
\usepackage{graphicx}
\usepackage{dcolumn}
\usepackage{bm}

\usepackage[mathlines]{lineno}
\usepackage[
  unicode=true,        
  colorlinks=true,     
  linkcolor=blue,      
  citecolor=blue,
  urlcolor=blue,      
  breaklinks=true      
]{hyperref}

\begin{document}


\title{Implementation of non-local arbitrary two-qubit controlled gates  via geometric quantum computation with Rydberg anti-blockade
}%

\author{Le-Jiang Yu}
\affiliation{Department of Physics, Zhejiang Normal University, Jinhua 321004, China}
\author{Jia Zheng}
\affiliation{Department of Physics, Zhejiang Normal University, Jinhua 321004, China}
\author{Kun Pu}
\affiliation{Department of Physics, Zhejiang Normal University, Jinhua 321004, China}
\author{Chao Gao}
\email{gaochao@zjnu.edu.cn}
\affiliation{Department of Physics, Zhejiang Normal University, Jinhua 321004, China}
\affiliation{Zhejiang Institute of Photoelectronics, Jinhua 321004, China}
\affiliation{Key Laboratory of Optical Information Detection and Display Technology of Zhejiang, Zhejiang Normal University, Jinhua, 321004, China}

\begin{abstract}
In the context of Rydberg anti-blockade, this paper proposes a new scheme for a high-fidelity controlled-unitary gate based on non-adiabatic holonomic quantum computation. 
Under specific detuning and interaction conditions, the scheme achieves a suitable evolution path for non-adiabatic holonomic quantum computation through reverse engineering of pulse parameters. This scheme can implement arbitrary two-qubit controlled gates with favorable robustness. We also evaluate the effects of spontaneous emission and laser intensity errors on the fidelity; numerical simulations show that the geometric gate scheme still exhibits excellent performance under these errors. Furthermore, we extend the local gates to non-local operations by combining entanglement transfer with auxiliary-atom measurements and classical feedforward, and demonstrate their application to interconversion among four-qubit entangled states (GHZ, cluster, and $W$ states). This finding indicates the potential applicability of our scheme to complex quantum information processing tasks.

\end{abstract}


\maketitle


\section{INTRODUCTION}

The field of quantum information merges quantum mechanics and information science, demonstrating immense potential in accelerating complex problem-solving, ensuring secure information transmission, and achieving high-precision detection~\cite{monroe2002quantum}. Among its branches, quantum computing has already broken through fundamental limitations of classical computing in certain specific areas, showcasing significant application prospects~\cite{shor1999polynomial,grover1997quantum,king2025beyond,benedetti2026provable}. The key to realizing quantum computing, however, hinges on the precise manipulation of quantum logic gates, which directly determines the execution efficiency and computational power of quantum algorithms. Currently, achieving high-fidelity, scalable quantum gate operations across different physical platforms has become a core challenge in advancing the practical application of quantum computing.

The design of quantum logic gate schemes must meet sufficiently high fidelity and robustness requirements. Geometric quantum computation (GQC) , utilizing geometric phases that depend only on evolution paths rather than dynamical details, inherently provides robustness against local noise~\cite{grover1997quantum,wilczek1984appearance,zanardi1999holonomic,niu2021error}. While early adiabatic approaches required long evolution times that increased susceptibility to decoherence, the development of non-adiabatic holonomic quantum
computation (NHQC) overcame these adiabatic limitations~\cite{zhang2023geometric, sonnerborn2024estimate, neef2025non,xu2017composite,zhao2018nonadiabatic,liang2024nonadiabatic,jin2025universal}. However, NHQC still needs to satisfy two conditions: (\text{i}) the cyclic evolution condition and (\text{ii}) the parallel transport condition~\cite{wu2022unselective}. Currently, researchers have conducted extensive studies on NHQC-based quantum gate schemes across different physical platforms such as superconducting circuits~\cite{li2020fast}, NV centers~\cite{zu2014experimental}, and nuclear magnetic resonance~\cite{zhu2002implementation}, gradually relaxing the restrictions of non-adiabatic holonomic conditions~\cite{li2021noncyclic,guo2020complete,guo2022multiple}. Among them, the NHQC+ scheme, in which condition (ii) is relaxed to only require the accumulated dynamic phase throughout the entire process to be zero, has been widely applied~\cite{liu2019plug}.

Currently, Rydberg atom platforms have emerged as a highly promising system in the field of quantum information processing, offering unique advantages such as strongly tunable interactions, long coherence times, and high-fidelity quantum operations~\cite{morgado2021quantum}. Leveraging the dipole-dipole interactions between Rydberg atoms, the Rydberg blockade effect suppresses simultaneous excitation of nearby atoms, providing a reliable scheme for constructing high-fidelity two-qubit quantum logic gates, which have been widely applied in quantum computing and simulation~\cite{beterov2020application,li2021multiple,guo2020optimized,liang2025error}. Conventional Rydberg-blockade gates utilize the interaction-induced energy shift to suppress the simultaneous excitation of two atoms. However, a finite blockade strength can result in residual double-excitation leakage and coherent gate errors. Moreover, implementing multi-qubit gates using the blockade effect requires complex temporal sequencing.~\cite{ming2024new}. 
These limitations have motivated the pursuit of different interaction regimes, with the Rydberg anti-blockade effect standing out as a robust and versatile candidate.

The Rydberg anti-blockade effect enables simultaneous excitation of multiple atoms into Rydberg states. 
Experimentally, by adjusting the detuning between the laser frequency and atomic transition frequency, resonant transitions can be achieved when specific conditions are met through compensation of the energy shifts induced by Rydberg interactions~\cite{su2017fast}. 
To date, NHQC-based two-qubit gate schemes utilizing the Rydberg anti-blockade effect have primarily focused on specific operations such as SWAP or iSWAP gates~\cite{wu2022unselective,xiao2024effective,li2024high,wu2021one}. 
Although a universal set of holonomic gates was recently demonstrated via an elegant dark-path scheme under off-resonant driving~\cite{jin2024geometric}, implementing high-fidelity arbitrary controlled-unitary (CU) gates specifically through the Rydberg anti-blockade mechanism remains an important and distinct challenge.


In this paper, we develop a comprehensive theoretical framework for implementing non-local arbitrary two-qubit CU gates by synergizing the Rydberg anti-blockade effect with NHQC+ dynamics.
Unlike previous anti-blockade schemes that often rely on specific dark-path evolutions or require the control atom to reside in the Rydberg state, our protocol achieves geometric gate operations through a reverse-engineering pulse design based on dynamical invariants, while keeping the Rydberg state outside the computational subspace to inherently suppress leakage errors~\cite{cong2022hardware}. 
Through rigorous numerical simulations, we demonstrate that the proposed geometric gates exhibit high fidelity and remarkable robustness against both spontaneous emission and laser intensity fluctuations under realistic experimental parameters. 
Furthermore, we extend the local gate operations to a non-local
regime through entanglement transfer and measurement-based
classical feedforward, providing a route to long-range qubit
manipulation within a single Rydberg array.
To highlight the practical utility of our scheme, we propose efficient quantum circuits for the mutual conversion between four-qubit GHZ, cluster, and $W$ states, thereby offering a versatile toolbox for multi-qubit entanglement resource management in large-scale quantum information processing.

\section{Rydberg antiblockade CU gate based on NHQC+ dynamics}
\subsection{Model and realization of the gates}
\begin{figure}[h]
\centering\includegraphics[scale=0.7]{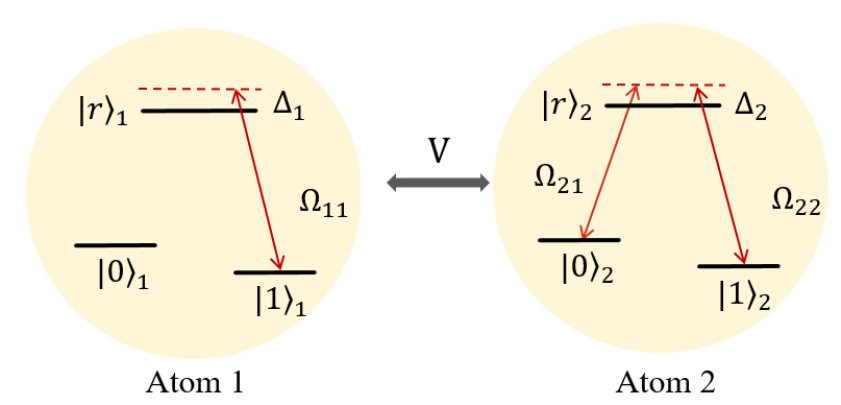}
\caption{Schematic diagram of three-level Rydberg atoms interacting with lasers.}
\label{fig1}
\end{figure}

 Now we demonstrate how to construct a two-qubit CU gate using NHQC+ dynamics. In our scheme, there are two Rydberg atoms, each consisting of three energy levels: two ground states \(|0\rangle\), \(|1\rangle\), and a Rydberg state \(|r\rangle\) (as shown in Fig. \ref{fig1}). For atom 1, the \(|0\rangle\) state remains idle, and the transition  \(\vert 1 \rangle_1 \leftrightarrow \vert r \rangle_1\)  is driven by a laser with Rabi frequency \(\Omega_{11}\) and blue detuning \(\Delta_1\). For atom 2, the \(|0\rangle\) and \(|1\rangle\) states are coupled to \(|r\rangle\) with Rabi frequencies \(\Omega_{21}\) and \(\Omega_{22}\), respectively, and with the same detuning \(\Delta_2\). Under the interaction picture, the Hamiltonian of the system can be written as
\begin{equation}\label{eq1}
\hat{H}=\hat{H}_{1}+\hat{H}_{2}+\hat{V}_{r r},
\end{equation}
where \(\hat{H}_1\) and \(\hat{H}_2\) are respectively the Hamiltonians of atom 1 and atom 2 (taking \(\hbar = 1\)):
\begin{align}
\hat{H}_{1}
&=e^{-i \Delta_{1} t}\Omega_{11}|1\rangle_{1}\langle r|+\text { H.c. },\nonumber\\
\hat{H}_{2}
&=e^{-i \Delta_{2} t}\left(\Omega_{21}|0\rangle_{2}\langle r|+\Omega_{22}|1\rangle_{2}\langle r|\right)+\text { H.c. },
\end{align}
and the third term in Eq.~(\ref{eq1}) describes the Rydberg interaction:
\begin{equation}
\hat{V}_{rr} = V \vert rr \rangle \langle rr \vert.
\end{equation}

In the two-atom basis, by applying the unitary operator \( \hat{U}=\exp \left[-i\left(\Delta_{2}-\Delta_{1}\right)t|r r\rangle\langle r r| \right]\), the system is shifted to the rotating frame, giving rise to
\begin{equation}\label{eq4}
\begin{aligned}
\hat{H}^{\prime} & =\hat{U}^{\dagger} \hat{H} \hat{U}+i \frac{d \hat{U}^{\dagger}}{d t} \hat{U} \\
& =\left[\Omega_{11} e^{-i \Delta_{1} t}(|10\rangle\langle r 0|+|11\rangle\langle r 1|)
    +\Omega_{11} e^{i\left(\Delta_{2}-2 \Delta_{1}\right) t}|1 r\rangle\langle r r|\right.\\
& +\Omega_{21} e^{-i \Delta_{2} t}(|0 r\rangle\langle 00|+|1 r\rangle\langle 10|)
    +\Omega_{21} e^{-i \Delta_{1} t}|r r\rangle\langle r 0|\\
& \left. +\Omega_{22} e^{-i \Delta_{2} t}(|0 r\rangle\langle 01|+|1 r\rangle\langle 11|)
    +\Omega_{22} e^{-i \Delta_{1} t}|r r\rangle\langle r 1| +\text{H.c.}\right] \\
& +\left(V+\Delta_{1}-\Delta_{2}\right)|r r\rangle\langle r r|.
\end{aligned}
\end{equation}
Under the  conditions \(\Delta_1,\Delta_2 \gg \Omega_{11},\Omega_{21},\Omega_{22}\) and \(\Delta_2 \gg \Delta_1\),we can obtain the effective Hamiltonian~(see Appendix~\ref{app})
\begin{equation}\label{eq5}
\hat{H}_\text{eff}  =\left(\Omega_{1}|rr\rangle\langle 11|+\Omega_{2}|rr\rangle\langle 10|+\text { H.c. }\right)+ \delta \vert rr \rangle \langle rr \vert ,
\end{equation}
where
\begin{equation}\label{eq6}
\Omega_{1}=-\frac{\Omega_{21} \Omega_{11}}{\Delta_{1}}, \Omega_{2}=-\frac{\Omega_{22} \Omega_{11}}{\Delta_{1}}. 
\end{equation}
Here, the Stark shift terms are omitted because they can be eliminated by using auxiliary pulses or levels. To realize the Rydberg antiblockade effect and resonantly couple the ground states \(|10\rangle\) and \(|11\rangle\) with the double Rydberg state \(|rr\rangle\), it is necessary to set \(\delta=0\), which can be achieved by setting specific values of \(V\). However, in the following, we adopt time-dependent Rabi frequencies, so the calculation of 
\(V\) is ultimately determined by the following expression,
\begin{equation}
V = \Delta_2 - \Delta_1 + \frac{|\Omega_{21}^{\text{max}}|^2 + |\Omega_{22}^{\text{max}}|^2}{\Delta_1} + \frac{|\Omega_{11}^{\text{max}}|^2}{\Delta_2 - 2\Delta_1}.
\end{equation}
The resulting off-peak residual detuning is retained in the numerical simulation.

To implement the geometric CU gate via Rydberg atoms, we set the driving field as \(\Omega_{1}=\Omega_{0} \sin \frac{\theta}{2} e^{i \varphi}, \Omega_{2}=-\Omega_{0} \cos \frac{\theta}{2} \).
Accordingly, the Hamiltonian becomes
\begin{equation}\label{eq7}
\hat{H}_\text{eff}^{\prime} = \Omega_0 |rr\rangle \langle B| + \text{H.c.}
\end{equation}
Here \(\left| B \right\rangle = \sin \frac{\theta}{2} e^{i \varphi} \left| 10 \right\rangle - \cos \frac{\theta}{2} \left| 11 \right\rangle\) is called bright state. 
The evolution of the effective Hamiltonian actually corresponds to Rabi oscillations between the bright state and the double Rydberg state \(\vert rr \rangle\). 
This Hamiltonian also admits a dark eigenstate \(\vert D \rangle = \cos \frac{\theta}{2} \vert 10 \rangle + \sin \frac{\theta}{2} e^{-i\varphi} \vert 11 \rangle\) with a zero eigenvalue, which is decoupled from the dynamics.

Next, we will show how to construct a two-qubit gate through reverse engineering using the dynamics of NHQC+. The system’s evolution under the effective Hamiltonian \(\hat{H}_\text{eff}^{\prime} \) is governed by the time-dependent Schrödinger equation:
\begin{equation}\label{eq8}
i \frac{\partial}{\partial t}|\psi(t)\rangle=\hat{H}_\text{eff}^{\prime}|\psi(t)\rangle.
\end{equation}
In  the computational subspaces \(|rr\rangle\) and \(|B\rangle\), by defining two time-dependent angles \(\alpha\) and \(\beta\), the state can be parameterized as follows:
\begin{equation}\label{eq9}
\left|\psi(t)\right\rangle=\cos \frac{\beta}{2}|r r\rangle+i e^{-i \alpha} \sin \frac{\beta}{2}|B\rangle.
\end{equation}
Simultaneously, the orthogonal evolved state \(\left|\psi_{\perp}(t)\right\rangle\) also satisfy Eq.~(\ref{eq8}), thus it can be parameterized as
\begin{equation}\label{eq10}
\left|\psi_{\perp}(t)\right\rangle=i e^{i \alpha} \sin \frac{\beta}{2}|r r\rangle+\cos \frac{\beta}{2}|B\rangle.
\end{equation}
To derive the pulse constraint equation determined by  \(\alpha\) and\(\beta\), we first substitute \(\Omega_{0}\) in Eq.~(\ref{eq7}) with \(\Omega_{0}=\Omega_{R}+i\Omega_{I}\). Then, based on Eqs.~\eqref{eq7}--\eqref{eq10}, we can obtain
\begin{equation}
\begin{array}{l}
\Omega_{R}=(\dot{\alpha} \sin \alpha \tan \beta-\dot{\beta} \cos \alpha) / 2, \\
\left.\Omega_{I}=(\dot{\alpha} \cos \alpha \tan \beta+\dot{\beta} \sin  \alpha\right) / 2.
\end{array}
\end{equation}
Based on the above equation, a suitable set of $\alpha$ and  $\beta$ can be chosen to implement the evolution of the quantum gate. To satisfy the cyclic evolution condition, it is also necessary to ensure the boundary condition \(\beta(0) = \beta(T)\) holds. To obtain a pure geometric phase, we adopt the setting method from Ref.~\cite{xiao2024effective} for 
\(\beta\) and \(\alpha\):
\begin{equation}\label{eq13}
\begin{array}{l}
\beta(t)=\pi \sin ^{2}(\pi t / T), \\
\alpha(t)=-\gamma \varepsilon(t)+4/3 \sin ^{3}\beta(t) , \\
\varepsilon(t)=\left\{\begin{array}{l}
0, t \in[0, T / 2), \\
1, t \in[T / 2, T],
\end{array}\right.
\end{array}
\end{equation}
such that upon evolving for one period \(T\), the dynamic phase is zero and the total phase is \(\gamma\). Therefore, the evolution operator in computational basis \(\{ |00\rangle, |01\rangle, |10\rangle, |11\rangle \}\) can be written as:
\begin{equation}\label{eq14}
\begin{aligned}
\hat{U}(\gamma, \theta, \varphi) & =|00\rangle\langle 00|+|01\rangle\langle 01|+|D\rangle\langle D|+e^{i \gamma}|B\rangle\langle B| \\
& =|0\rangle\langle 0| \otimes \hat{I}+|1\rangle\langle 1| \otimes \hat{u} ,
\end{aligned}
\end{equation}
where
\begin{equation}
\begin{aligned}
\hat{u} & =e^{i \frac{\gamma}{2}}\left(\begin{array}{cc}
\cos \frac{\gamma}{2}-i \cos \theta \sin \frac{\gamma}{2} & -i e^{i \varphi} \sin \theta \sin \frac{\gamma}{2} \\
-i e^{-i \varphi} \sin \theta \sin \frac{\gamma}{2} & \cos \frac{\gamma}{2}+i \cos \theta \sin \frac{\gamma}{2}
\end{array}\right) \\
& =e^{i \frac{\gamma}{2}} e^{-i \frac{\gamma}{2} \mathbf{n} \cdot \sigma}.
\end{aligned}
\end{equation}
Here, \(\sigma = (\sigma_x, \sigma_y, \sigma_z)\) denotes the standard Pauli matrices, \(\mathbf{n} = (\sin \theta \cos \varphi, -\sin \theta \sin \varphi, \cos \theta)\) denotes a unit vector.
The parameters \(\beta\) and \(\varphi\) are adjusted independently to implement an arbitrary two-qubit NHQC+ controlled gate, in which \(\hat{u}\) represents any single-qubit operation, and we can implement a CU gate by adjusting the values of the laser parameters \(\gamma\), \(\theta\) and \(\varphi\). For instance, we choose \(\gamma=\pi\), \(\theta=\frac{\pi}{2}\), \(\varphi=0\) for the CNOT gate; \(\gamma=\pi\), \(\theta=0\), \(\varphi=0\) for the CZ gate; and \(\gamma=\pi\), \(\theta=\frac{\pi}{4}\), \(\varphi=0\) for the CH gate.

It is worth noting that while the pioneering work in Ref.~\cite{jin2024geometric} elegantly achieves robust holonomic gates through a dark-path manifold under off-resonant driving, our approach leverages the Rydberg anti-blockade effect combined with the reverse engineering of pulse parameters. In our scheme, the singly excited Rydberg states are suppressed by large detunings, whereas the doubly excited state \(|rr\rangle\) is transiently populated during the gate evolution. In the ideal effective model, cyclic evolution returns the population to the computational subspace at the end of the gate.

\subsection{Robustness against  spontaneous emission and laser intensity errors}

Next, we will use numerical simulation to verify the feasibility of the logic gate scheme.
In the ideal case, for  the general initial state \(|\psi(0)\rangle=\left(\cos \theta|0\rangle_{1}+\sin \theta|1\rangle_{1}\right) \otimes\left(\cos \eta|0\rangle_{2}+\sin \eta|1\rangle_{2}\right)\), a quantum logic gate \(U\) can precisely evolve the state into the ideal target state :
\begin{equation}
|\psi_{\text{ideal}}\rangle = U |\psi(0)\rangle.
\end{equation}
However, in an open quantum system, the evolution is inevitably affected by noise, decoherence, and other environmental factors, causing the actual output state, described by the density matrix \(\rho\), to deviate from the ideal case. In this case, the fidelity of the quantum state is defined as:
\begin{equation}\label{eq16}
    F=\frac{1}{(2 \pi)^{2}} \int_{0}^{2 \pi} \int_{0}^{2 \pi}\langle\psi_{\text{ideal}}| \rho(t)|\psi_{\text{ideal}}\rangle d \theta d \eta,
\end{equation}
The dynamical evolution of an open system is controlled by the Lindblad master equation,
\begin{equation}
\dot{\rho}(t)=i[\rho(t), H(t)]+\sum_{j} \sum_{k}\left[L_{j}^{k} \rho L_{j}^{k\dagger}-\frac{1}{2}\left(L_{j}^{k\dagger} L_{j}^{k} \rho+\rho L_{j}^{k\dagger} L_{j}^{k}\right)\right],
\end{equation}
where \(H(t)\) represents the total Hamiltonian of the system, and \(\rho(t)\) denotes the density matrix of the system state. For Rydberg atoms, the spontaneous decay of Rydberg states constitutes the primary factor affecting system fidelity, described by Lindblad operators \(L_j^k = \sqrt{{\Gamma/2}} | k \rangle_j \langle r |\), where \(k = 0, 1\) denotes an arbitrary initial state, \(j\) is the atom index, and \(\Gamma\) is the decay rate from the Rydberg state to the ground state.

\begin{figure}[t]
\centering\includegraphics[scale=0.45]{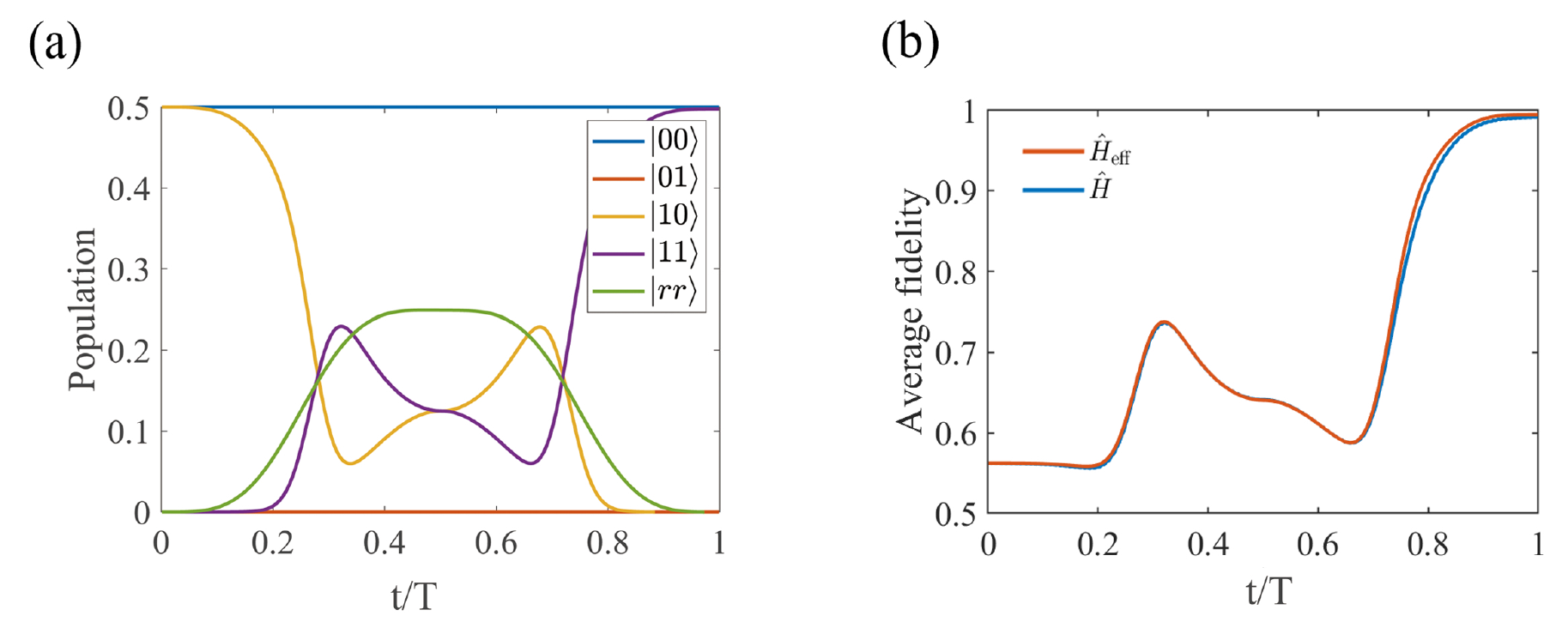}
\caption{(a) The population dynamics in the evolution of the CNOT gate. (b) The  fidelity of the CNOT gate changes over time, and the final  fidelity obtained is 0.995.}
\label{fig2}
\end{figure}

In our scheme, we need to select an appropriate detuning to satisfy the prerequisite conditions of the effective Hamiltonian. 
This affects the Rabi frequency of the laser, the gate evolution time, and the interaction strength between Rydberg atoms. 
Specifically, shorter gate evolution times lead to higher gate fidelity, but this requires larger Rabi frequencies, which in turn result in greater detuning and stronger Rydberg interaction strengths. Next we use the CNOT gate as an example to illustrate parameter selection in our scheme. 
For the Rabi frequency of the laser, we adopted a typical experimental value of \(\Omega_{11}= 2\pi \times 1.7 \text{ MHz}\) and \(\quad \Omega_{21} = \Omega_{22} = 2\pi \times 1.2 \text{ MHz}\), with corresponding detunings of \(\Delta_1 = 2\pi \times 16.6 \text{ MHz}\) and \(\Delta_2 = 2\pi \times 694.5 \text{ MHz}\), respectively. The Rydberg interaction strength is \(V=2\pi \times 684.6 \text{ MHz}\). In this paper, we assume the Rydberg atoms to be $^{87}\mathrm{Rb}$ atoms, with the Rydberg state set as \(|70S_{1/2}\rangle\). According to Ref.~\cite{beterov2009quasiclassical}, at 0K, the lifetime \(\tau\) of the Rydberg state is 416 \(\mu s\), and the decay rate \(\Gamma=1/\tau\approx2.4 \text{kHz}\). 
Under these parameters,  the calculated  fidelity of the CNOT gate is shown in  Fig.~\ref{fig2}(b), where the gate operation time is  11 \(\mu s\). The fidelities driven by the original Hamiltonian $H$ and the effective Hamiltonian \(H_\text{eff}\) are nearly identical. Specifically, the  fidelities of the CNOT gate calculated from 
are 0.992 and 0.995, respectively, which demonstrates the validity of our original Hamiltonian.

\begin{figure}[h]
\centering\includegraphics[scale=0.27]{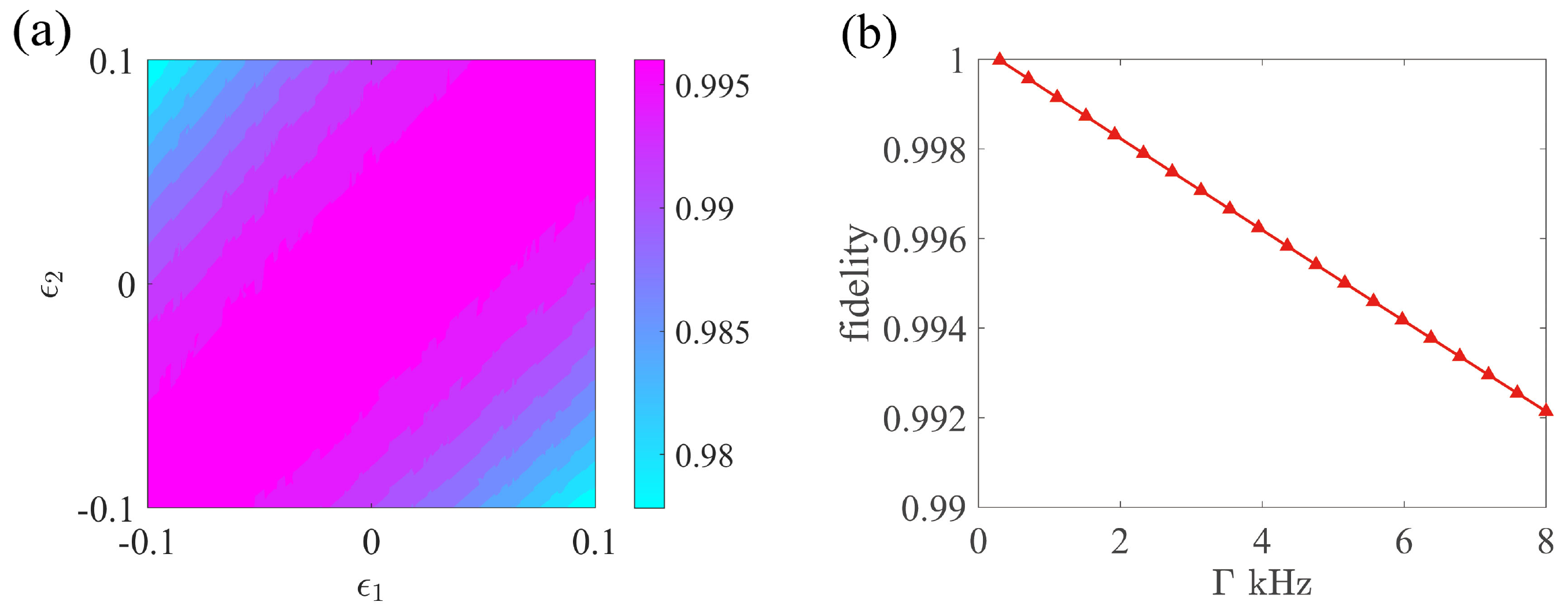}
\caption{(a) Schematic diagram of fidelity under laser error. (b) Fidelity under different decay rates.}
\label{fig3}
\end{figure}

 In addition, due to the instability of the laser, the Rabi frequency may admit certain errors. We consider the driving pulses \(\Omega_1\) and \(\Omega_2\) varying within the ranges \((1+\epsilon_1)\Omega_1\) and \((1+\epsilon_2)\Omega_2\), respectively, where the error fractions \(\epsilon_1, \epsilon_2\in[-0.1,0.1]\). In Fig.\ref{fig3}(a), we plot the fidelity variation caused by laser errors, it can be seen from the figure that Synchronous fluctuations largely preserve the ratio \(\Omega_1/\Omega_2\) and thus the geometric evolution path, whereas asynchronous fluctuations distort this ratio and cause larger gate errors. Fig.\ref{fig3}(b) shows the fidelity under different decay rates, with \(\Gamma\) ranging from 0.2 to 5 kHz. Overall, the fidelity under parameter fluctuations demonstrates that our scheme is indeed robust.

\section{Non-local CU gate protocol}
Quantum entanglement constitutes a fundamental resource in quantum computing, which necessitates the development of reliable quantum gates to facilitate operations on entangled atomic systems.
Nevertheless, many quantum computing architectures demand long-distance entanglement between qubits. 
In these scenarios, conventional Rydberg-mediated quantum gates are inapplicable, as they impose stringent constraints on the spatial proximity of the atoms.
Although optical tweezers can be used to move atoms in Rydberg arrays, atom transport increases susceptibility to decoherence ~\cite{berthusen2025toward}. Therefore, the construction of non-local quantum gates is essential. 

 Over the past few decades, researchers have proposed several schemes for implementing non-local CNOT gates ~\cite{han2022scheme,liu2019heralded,gottesman1999demonstrating,huang2004experimental,Eisert2000}. In particular, Eisert et al.\cite{Eisert2000} proved that their protocol achieves optimal resource consumption when only local operations and classical communication are allowed; therefore, inspired by their work, we extend the previously proposed NHQC+ CU gate to a nonlocal CU gate. The geometric character pertains to the local two-qubit gates implemented through the NHQC+ dynamics.
The complete non-local protocol additionally involves
entanglement transfer, single-qubit mapping pulses,
auxiliary-atom measurements, and classical feedforward,
which are not generally geometric operations.

\begin{figure}[h]
\centering\includegraphics[scale=0.5]{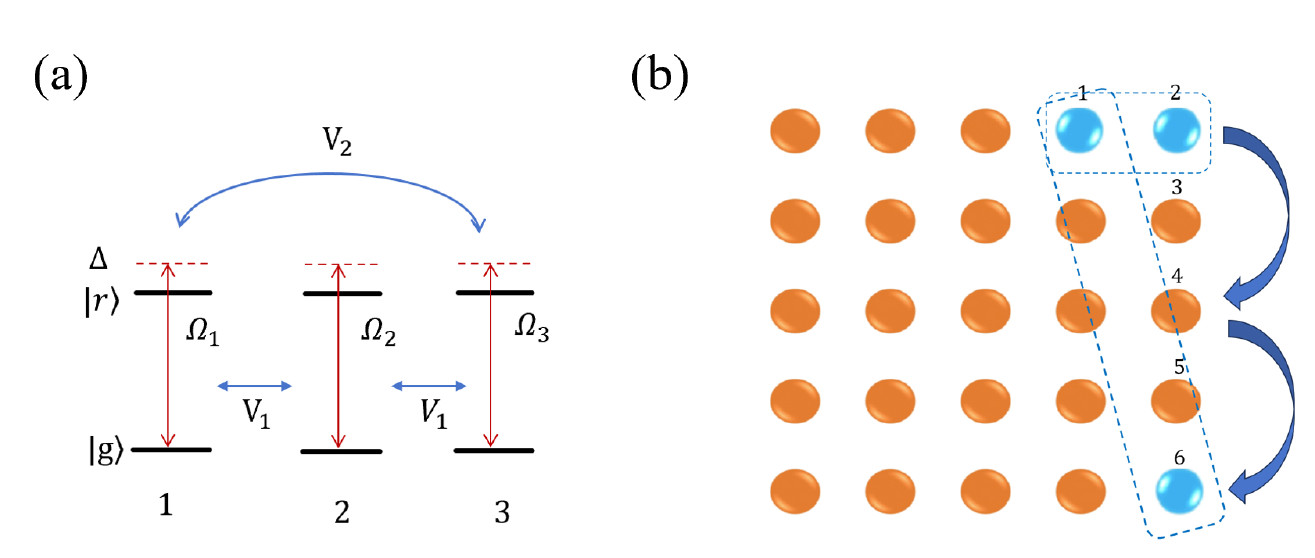}
\caption{(a) Schematics of interaction between three Rydberg atoms and lasers. (b) Diagram illustrating the process of entanglement transfer.}
\label{fig5}
\end{figure}
In non-local quantum gates, the entanglement channel should also be long-range. Wu et al.~\cite{wu2025perfect} proposed a technique for Bell state transfer utilizing the anti-blockade effect. Applying this technique to non-local quantum gate protocols can enhance the flexibility of our scheme. 
Considering a three-atom two-level system of Rydberg atoms as shown in Fig.~\ref{fig5}(a), when the antiblockade condition 
$\Delta = V_1, \; V_{1,2} \gg \Omega_{1,2,3}$ is satisfied, the three-atom system can be described by a five-site artificial-lattice Hamiltonian,
\begin{equation}\label{eq22}
\begin{aligned}
\hat{H}= & J_{1}|r g g\rangle\langle r r g|+J_{2}|r r g\rangle\langle g r g|+ \\
& J_{3}|g r g\rangle\langle g r r|+J_{1}|g r r\rangle\langle g g r|+\text{H.c.},
\end{aligned}
\end{equation}
Thus, this Hamiltonian enables the transfer of entanglement from adjacent atoms 1 and 2 to atoms 1 and 6 (as shown in Fig.~\ref{fig5}(b)). First, the six atoms are prepared in the following state:
\begin{equation}
|\psi(t=0)\rangle=\frac{|g r\rangle_{12}+|r g\rangle_{12}}{\sqrt{2}} \otimes|g g g g\rangle_{3456}.    
\end{equation}
Next, three laser beams with Rabi frequencies satisfying  Eq.~(\ref{D6}) are applied to atoms 2, 3, and 4, respectively. When the evolution time reaches $t=\pi/\Omega$, we find that the entire system evolves to
\begin{equation}
|\psi(t = \frac{\pi}{\Omega})\rangle=\frac{|g r\rangle_{14}+|r g\rangle_{14}}{\sqrt{2}} \otimes|g g g g\rangle_{2356}.    
\end{equation}
This process enables the transfer of the Bell state from atoms 1 and 2 to atoms 1 and 4. When the laser satisfying  Eq.~(\ref{D6}) are applied to atoms 4, 5, and 6, respectively, and after another evolution period of  $t=\pi/\Omega$, the final state of the entire system is calculated to be
\begin{equation}
|\psi(t = \frac{2\pi}{\Omega})\rangle=\frac{|g r\rangle_{16}+|r g\rangle_{16}}{\sqrt{2}} \otimes|g g g g\rangle_{2345}.    
\end{equation}
In this way, entanglement is successfully transferred from atoms 1 and 2 to atoms 1 and 6. This process can be repeated to propagate the entanglement even further. Ref.~\cite{wu2025perfect} quantifies the maximum number of such steps achievable with specific experimental parameters, proving that this approach is viable for large-scale quantum information transfer within the lifetime of the Rydberg state.

We now combine the distributed entangled channel with the local NHQC+ gate introduced in Sec. II. The choice of the entangled channel (\(\frac{1}{\sqrt{2}}(|r1\rangle+|1r\rangle)\)) is consistent with the teleportation protocol of Wu et al.~\cite{wu2013quantum}, and the same preparation method can be adopted. Once the entanglement is prepared  and transferred to the designated locations, we assume the control qubit carries information $\alpha|0\rangle+\beta|1\rangle$  and is labeled as atom 1, while the target qubit is $\delta|0\rangle+\gamma|1\rangle$ and labeled as atom 4. The two entangled atoms are labeled as atoms 2 and 3. Furthermore, it is required that atom 2 is within the Rydberg interaction radius of atom 1, and atom 3 is within the Rydberg interaction radius of atom 4. Under these conditions, the state of the entire system is 
\begin{equation}\label{25}
|\Phi_{1}\rangle=(\alpha|0\rangle+\beta|1\rangle)_1\otimes\frac{1}{\sqrt{2}}(|r1\rangle+|1r\rangle)_{23}\otimes(\delta|0\rangle+\gamma|1\rangle)_4.
\end{equation}

To connect the auxiliary encoding to the ground-state gates,
we implement the local mapping $R_k$, for $k=2,3$, using a
phase-calibrated resonant $\pi$ pulse on the
$|0\rangle_k\leftrightarrow|r\rangle_k$ transition, leaving
$|1\rangle_k$ unaffected.
The pulse phase is chosen such that
\begin{equation}
 R_k|r\rangle_k=|0\rangle_k,
 \qquad
 R_k|1\rangle_k=|1\rangle_k.
\label{eq:ncu_mapping_pulse}
\end{equation}
The relative phase between these two components is preserved.
During each mapping pulse, the other atom in the same local
pair occupies the ground-state manifold.

The gate sequence begins with $R_2$, followed by the local
$\mathrm{CNOT}_{1\rightarrow2}$. The latter is obtained from the
NHQC+ construction in Sec.~II .
The resulting four-atom state is
\begin{equation}
\begin{aligned}
 |\Phi_2\rangle=\frac{1}{\sqrt{2}}(\alpha|001\rangle+\alpha|01r\rangle+\beta|111\rangle+\beta|10r\rangle)_{123}\otimes(\delta|0\rangle+\gamma|1\rangle)_4.
\end{aligned}
\label{eq:ncu_after_alice_cnot}
\end{equation}
Atom $2$ is then measured in the computational basis, and the
one-bit outcome $m\in\{0,1\}$ is recorded by the classical
controller. The normalized conditional
state of atoms $1$, $3$, and $4$ is
\begin{equation}
 |\Phi_m\rangle=
 \begin{cases}
 (\alpha|01\rangle+\beta|1r\rangle)_{13}(\delta|0\rangle+\gamma|1\rangle)_4
 , & m=0,\\[1mm]
 (\alpha|0r\rangle+\beta|11\rangle)_{13}(\delta|0\rangle+\gamma|1\rangle)_4, & m=1.
 \end{cases}
\label{eq:ncu_first_measurement}
\end{equation}
The mapping $R_3$ is applied to atom $3$, followed by the
measurement-dependent correction $X_3^{\,1-m}$ on its ground-state
qubit. The controller therefore triggers an $X_3$ pulse only when
$m=0$. Both outcomes then yield
\begin{equation}
 |\Phi_3\rangle
 =(\alpha|00\rangle+\beta|11\rangle)_{13}(\delta|0\rangle+\gamma|1\rangle)_4.
\label{eq:ncu_correlated_control}
\end{equation}
The ground-state single-qubit operations are implemented by
individually addressed pulses. Here
$X_k=|0\rangle_k\langle1|+|1\rangle_k\langle0|$
and $Z_k=|0\rangle_k\langle0|-|1\rangle_k\langle1|$.
The $X_3$ correction is realized by a resonant microwave or effective
two-photon Raman $\pi$ pulse on the
$|0\rangle_3\leftrightarrow|1\rangle_3$ transition, with its
phase chosen for rotation about the $x$ axis \cite{jones2007fast}.
The $Z_1$ correction used below is realized by accumulating a
relative phase of $\pi$ between $|0\rangle_1$ and $|1\rangle_1$
using a differential ac Stark shift \cite{levine2019parallel}.
The Hadamard operation $H_3$ is implemented by a resonant
$\pi/2$ pulse on the ground-state transition with the appropriate
phase correction: a $Z_3$ phase operation, produced by a differential
ac Stark shift, is applied first, followed by a $\pi/2$ pulse whose
phase is chosen for rotation about the $y$ axis \cite{silvrio2022pulser}. 

The local NHQC+ controlled-unitary gate
$U_{\mathrm{CU}}^{(3,4)}$
is subsequently applied with atom $3$ as the control and atom $4$
as the target.
For the single-qubit operation $u=u(\gamma,\theta,\phi)$
introduced in Sec.~II, this gives
\begin{equation}
 |\Phi_4\rangle
 =\alpha|00\rangle_{13}(\delta|0\rangle+\gamma|1\rangle)_4+\beta|11\rangle_{13}u_4(\delta|0\rangle+\gamma|1\rangle)_4.
\label{eq:ncu_second_measurement}
\end{equation}
Next, $H_3$ is applied using the pulse sequence specified
above, and atom $3$ is measured in the computational basis.
For outcome $n\in\{0,1\}$, the normalized state of the two
data atoms becomes
\begin{equation}
 |\Phi_n\rangle
 =\alpha|0\rangle_1(\delta|0\rangle+\gamma|1\rangle)_4
  +(-1)^n\beta|1\rangle_1u_4(\delta|0\rangle+\gamma|1\rangle)_4.
\label{eq:ncu_second_measurement}
\end{equation}
The outcome $n$ is recorded by the same classical controller,
which triggers a $Z_1$ correction on atom $1$ only when $n=1$,
thereby implementing $Z_1^n$. The final state is therefore
\begin{equation}
|\Phi_{\mathrm{out}}\rangle
 =CU_{(1,4)}(\alpha|0\rangle+\beta|1\rangle)_1(\delta|0\rangle+\gamma|1\rangle)_4.
\label{eq:ncu_target_output}
\end{equation}
Both data atoms remain unmeasured, and the output control and target
qubits remain at their original locations. All four measurement
branches are accepted and corrected, so the ideal protocol is
deterministic.

\section{Application — 4-qubit entangled state transformation}
In the field of quantum information processing, different types of entangled states play distinct roles, and the transformation between entangled states can further enhance the efficiency of quantum computing~\cite{jin2025preparing,lami2026asymptotic,jiang2026one}. Recent theoretical breakthroughs have been achieved in three-qubit implementations using Rydberg atom arrays ~\cite{shao2023high,haase2021conversion,wang2024deterministic}. However, these schemes rely on designing specific Hamiltonians and impose strict requirements on the spatial arrangement and distances of the three atoms, which significantly limit their applicability. Moreover, extending these schemes to a larger number of qubits is challenging, as it requires extensive computational resources to derive feasible effective Hamiltonians. 

\begin{figure}[h]
\centering\includegraphics[scale=0.5]{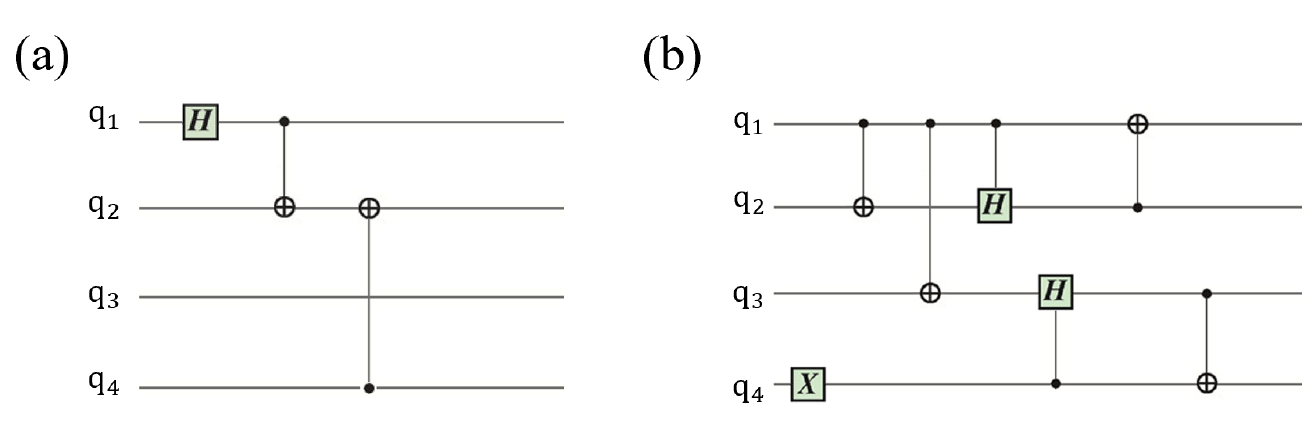}
\caption{Quantum circuits for entanglement conversion: (a) $GHZ$ to cluster state, (b) $GHZ$ to $W$ state. $q_i(i=1,2,3,4)$ denotes the qubit index.}
\label{fig6}
\end{figure}

As a highly entangled resource, cluster states possess exceptional robustness and persistence, making them promising candidates for a wide range of quantum tasks~\cite{yu2024generation}. 
However, a cluster state requires at least four qubits to form. 
To date, the mutual transformation between cluster states and other four-qubit entangled states remains largely unexplored. In this paper, we demonstrate that such transformations can be systematically and efficiently realized using our proposed quantum gates.
Before introducing the quantum circuit we designed for entangled state conversion, we first present the basic forms of the employed four-qubit entangled states:
\begin{equation}
\begin{aligned}
|GHZ\rangle &=   \frac{1}{\sqrt{2}}\left(|0000\rangle+|1111\rangle\right),\\
|\text{cluster}\rangle &=   \frac{1}{2}\left(|0000\rangle+|1100\rangle+|0011\rangle-|1111\rangle\right),\\
|W\rangle &=   \frac{1}{2}\left(|0001\rangle+|0010\rangle+|0100\rangle+|1000\rangle\right).
\end{aligned}
\end{equation}

\begin{figure}[h]
\centering\includegraphics[scale=0.45]{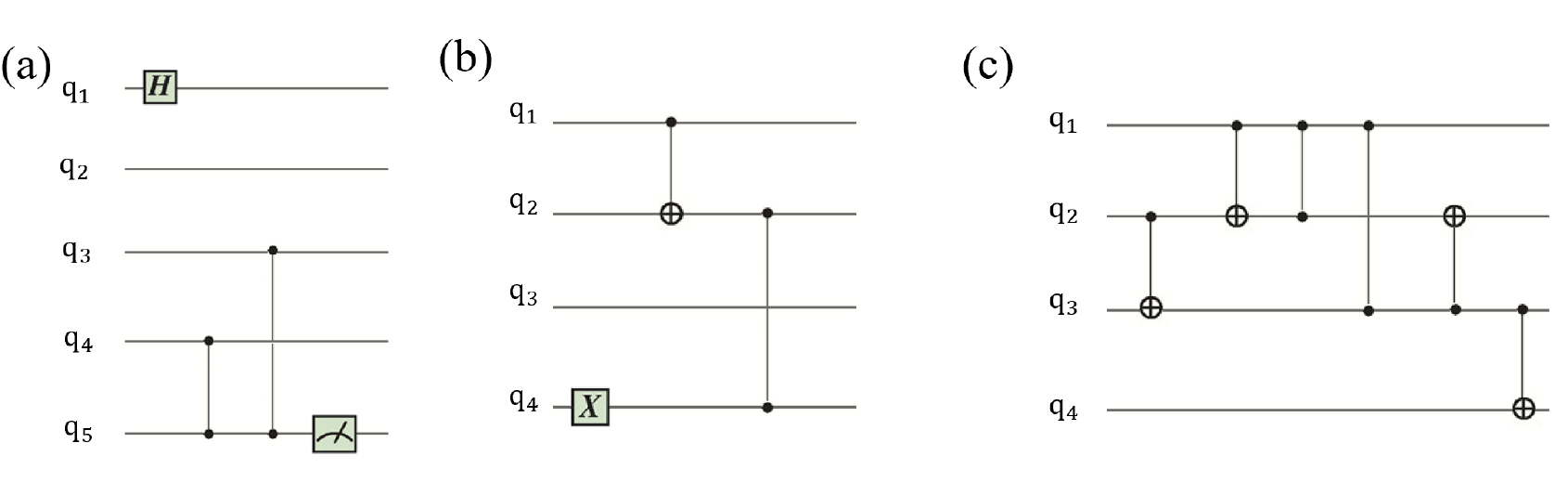}
\caption{(a) The entanglement conversion circuit from $W$ state to cluster state before measurement. Qubit $q_5$ is the ancilla, which is measured in the X-basis. (b) Feed-forward circuit for ancila qubit measurement result 1. (c) Feed-forward circuit for ancila qubit measurement result 0.}
\label{fig7}
\end{figure}

The conversion from a 4-qubit $GHZ$ state to a cluster state requires only one single-qubit Hadamard gate and two CNOT gates. Fig.~\ref{fig6}(a) illustrates one possible scheme, as the position of the single-qubit Hadamard operation in the first step can be chosen arbitrarily. To revert from the cluster state back to the $GHZ$ state, one simply needs to apply the inverse of this circuit. For the conversion from a 4-qubit $GHZ$ state to a $W$ state, the quantum circuit is shown in Fig.~\ref{fig6}(b). The process begins with the application of a single-qubit $X$ gate, followed by two CNOT gates, then two CH gates, and finally two more CNOT gates. Similarly, this circuit is not unique—the position of the $X$ gate can be changed, which in turn alters the control and target qubits of the subsequent controlled gates. For the conversion from the $W$ state to the cluster state, one can concatenate the inverse of the $GHZ$-to-$W$
 circuit in Fig.~\ref{fig6} with the $GHZ$-to-cluster-state circuit, thereby realizing the conversion from the $W$ state to the cluster state, with a total of eight two-qubit gates required. Here, we consider the measurement-assisted construction shown in Fig.~\ref{fig7}. With the ancilla initialized in \(|+\rangle\), the two equiprobable measurement outcomes require four and eight two-qubit gates, respectively, including the premeasurement operations. This construction therefore reduces the expected two-qubit gate count to six compared with the above ancilla-free circuit, at the cost of one ancillary qubit, measurement, and classical feedforward.

\section{Conclusion}
In summary, we have proposed a robust and flexible scheme for constructing arbitrary two-qubit controlled gates based on non-adiabatic holonomic quantum computation in the context of Rydberg anti-blockade.
By reverse-engineering the pulse parameters, the gate is designed to return the population to the computational subspace after an ideal evolution cycle, while allowing transient occupation of the doubly excited Rydberg state. Numerical simulations confirm that the scheme maintains high fidelity and excellent performance even in the presence of spontaneous emission and laser intensity fluctuations.
It is worth noting that the pioneering work in Ref.~\cite{jin2024geometric} elegantly achieves robust gates via a dark-path manifold under off-resonant driving, and our approach provides a complementary paradigm utilizing the Rydberg anti-blockade mechanism. 
A distinctive feature of our work is the extension of these local operations to non-local regimes by combining entanglement transfer
with auxiliary-atom measurements and classical feedforward.
We have demonstrated that this non-local extension enables the flexible mutual conversion among four-qubit entangled states, including GHZ, cluster, and $W$ states.
These protocols require only basic quantum gate operations, thereby avoiding the design of complex many-body Hamiltonians and significantly enhancing the flexibility and operability of entanglement resource management. 
This study not only extends the application of geometric quantum computation in Rydberg atom platforms but also provides valuable insights for advancing large-scale quantum information processing.

\begin{acknowledgments}
We thank Prof. J. Jing and Prof. G.-Q. Li for valuable comments and suggestions on the manuscript. This work was supported by Quantum Science and Technology-National Science and Technology Major Project~(2025ZD0300400), and the National Natural Science Foundation of China (12522413, 12474493).

\end{acknowledgments}

\section*{disclosures}
The authors declare that there are no conflicts of interest related to this article.

\appendix

\section{
DERIVATION OF THE EFFECTIVE HAMILTONIAN}\label{app}

In this appendix, we demonstrate the derivation of Eqs.~\eqref{eq5} and \eqref{eq6}. First, we rewrite Eq.~(\ref{eq4}) into two parts:
\begin{equation}
\hat{H}^{\prime}=\hat{H}_{0}+\hat{H}_{I},
\end{equation}
where
\begin{equation}
\begin{aligned}
\hat{H}_{0} &=\Omega_{11} e^{-i \Delta_{1} t}(|10\rangle\langle r 0|+|11\rangle\langle r 1|) \\
& +\Omega_{21} e^{-i \Delta_{2} t}(|0 r\rangle\langle 00|+|1 r\rangle\langle 10|) \\
& +\Omega_{22} e^{-i \Delta_{2} t}(|0 r\rangle\langle 01|+|1 r\rangle\langle 11|) \\
& +\Omega_{21} e^{-i \Delta_{1} t}|r r\rangle\langle r 0|+\Omega_{22} e^{-i \Delta_{1} t}|r r\rangle\langle r 1|) \\
& +\Omega_{11} e^{i\left(\Delta_{2}-2 \Delta_{1}\right) t}|1 r\rangle\langle r r|+\text{H.c.} ,
\end{aligned}
\end{equation}
and
\begin{equation}
\begin{aligned}
\hat{H}_{I}=\left(V+\Delta_{1}-\Delta_{2}\right)|r r\rangle\langle r r|.
\end{aligned}
\end{equation}

According to the effective Hamiltonian theory, the Hamiltonian of \(\hat{H}_{0}\) is written in the following form:
\begin{equation}
\hat{H}_{0}=\sum_{n=1,2,3} \hat{h}_{n}^{\dagger} e^{i \Delta_{n} t}+\hat{h}_{n} e^{-i \Delta_{n} t},
\end{equation}
in which \(\hat{h}_{1}\), \(\hat{h}_{2}\), and \(\hat{h}_{3}\) are of the following forms:
\begin{equation}\label{A5}
\begin{aligned}
\hat{h}_1 &= \Omega_{11} (|10\rangle\langle r0| + |11\rangle\langle r1|) + \Omega_{21}|rr\rangle\langle r0| + \Omega_{22}|rr\rangle\langle r1|, \\
\hat{h}_2 &= \Omega_{21} (|0r\rangle\langle 00| + |1r\rangle\langle 10|) + \Omega_{22}(|0r\rangle\langle 01| + |1r\rangle\langle 11|), \\
\hat{h}_3 &= \Omega_{11} |1r\rangle\langle rr|.
\end{aligned}
\end{equation}
Under the  conditions \(\Delta_1,\Delta_2 \gg \Omega_{11},\Omega_{21},\Omega_{22}\), after removing the high-frequency oscillation terms, \(\hat{H}_{0}\) becomes a simplified form:
\begin{equation}\label{A6}
\hat{H}_{0,\text{eff}}=\frac{1}{\Delta_{1}}\left[\hat{h}_{1}^{\dagger}, \hat{h}_{1}\right]+\frac{1}{\Delta_{2}}\left[\hat{h}_{2}^{\dagger}, \hat{h}_{2}\right]+\frac{1}{\Delta_{2}-2\Delta_{1}}\left[\hat{h}_{3}^{\dagger}, \hat{h}_{3}\right] .
\end{equation}
By inserting Eq.~(\ref{A5}) into ~(\ref{A6}) the effective Hamiltonian
can be changed to
\begin{equation}
\begin{aligned}
\hat{H}_\text{eff} & =\left[\frac{\Omega_{21}^{2}}{\Delta_{2}}|00\rangle\langle 00|+\left(\frac{\Omega_{22}^{2}}{\Delta_{2}}-\frac{\Omega_{11}^{2}}{\Delta_{1}}\right)|11\rangle\langle 11|+\frac{\Omega_{22}^{2}}{\Delta_{2}}|01\rangle\langle 01|+\left(\frac{\Omega_{21}^{2}}{\Delta_{2}}-\frac{\Omega_{11}^{2}}{\Delta_{1}}\right)|10\rangle\langle 10|\right. \\
& +\frac{\Omega_{11}^{2}+\Omega_{21}^{2}}{\Delta_{1}}(|r 0\rangle\langle r 0|+|r 1\rangle\langle r 1|)-\frac{\Omega_{21}^{2}+\Omega_{22}^{2}}{\Delta_{2}}(|0 r\rangle\langle 0 r|+|1 r\rangle\langle 1 r|)-\frac{\Omega_{21} \Omega_{11}}{\Delta_{1}}|r r\rangle\langle 10| \\
& \left.-\frac{\Omega_{22} \Omega_{11}}{\Delta_{1}}|r r\rangle\langle 11|+\frac{\Omega_{22}^{2}+\Omega_{21}^{2}}{\Delta_{2}}(|01\rangle\langle 00|+|11\rangle\langle 10|)+\text { H.c. }\right] \\
& +\left(\mathrm{V}+\Delta_{1}-\Delta_{2}-\frac{\Omega_{21}^{2}+\Omega_{22}^{2}}{\Delta_{1}}-\frac{\Omega_{11}^{2}}{\Delta_{2}-2 \Delta_{1}}\right)|r r\rangle\langle r r|.
\end{aligned}
\end{equation}
By omitting the Stark shift terms and setting \(\delta=0\), we obtain the following equation,

\begin{equation}
\begin{aligned}
\hat{H}_\text{eff}  =-\frac{\Omega_{21} \Omega_{11}}{\Delta_{1}}|r r\rangle\langle 10|  \left.-\frac{\Omega_{22} \Omega_{11}}{\Delta_{1}}|r r\rangle\langle 11|+\frac{\Omega_{22}^{2}+\Omega_{21}^{2}}{\Delta_{2}}(|01\rangle\langle 00|+|11\rangle\langle 10|)+\text { H.c. }\right
.
\end{aligned}
\end{equation}
If the condition \(\Delta_2 \gg \Delta_1\) is satisfied, the undesired Raman couplings can be neglected, and Eq.~(\ref{eq7}) is obtained.

\section{
SOLUTION OF THE RABI FREQUENCY OF
CLASSICAL FIELDS}\label{app2}
According to the designed control fields \(\Omega_x(t)\) and \(\Omega_y(t)\),
we can inversely solve the Rabi frequency of the classical field
as follows:
\begin{equation}
\begin{aligned}
\Omega_{11}(t)&=\sqrt{|\Omega_0(t)|\cdot\Delta_1},\\[6pt]
\Omega_{21}(t)&=
-\frac{\left(\Omega_0(t)\sin\frac{\theta}{2}e^{i\phi}\right)\cdot\Delta_1}
{\Omega_{11}(t)},\\[6pt]
\Omega_{22}(t)&=
-\frac{\left(-\Omega_0(t)\cos\frac{\theta}{2}\right)\cdot\Delta_1}
{\Omega_{11}(t)}.
\end{aligned}
\end{equation}

\section{CALCULATION OF THE GEOMETRIC PHASE}
To realize NHQC, it is only necessary
to eliminate the dynamical phase 
obtained in one evolution cycle as
\begin{equation}
\gamma_d = - \int_0^T \langle \phi_l(t) | H(t) | \phi_l(t) \rangle \, dt = 0.
\end{equation}
Then, the total phase is described by the geometric phase,
\begin{equation}
\gamma_g= \int_0^T \langle \phi_l(t) | i \frac{\partial}{\partial t} | \phi_l(t) \rangle \, dt,
\end{equation}
According to Eqs.~(\ref{eq9}) and ~(\ref{eq10}), the dynamical phase and the geometric phase can be obtained in the following forms,
\begin{equation}
\dot{\gamma_d} = \mp \frac{\dot{\alpha} \sin^2 \beta}{2 \cos \beta}, \quad \dot{\gamma_g} = \pm \dot{\alpha} \sin^2 \frac{\beta}{2}.
\end{equation}
Substituting Eq.~(\ref{eq13}) and after calculation, we obtain \(\gamma_d=0\)
, meaning that the evolution of the CU gate is purely geometric, this is consistent with the derivation in Ref.~\cite{wu2022unselective}.

\section{DERIVATION OF THE RYDBERG ANTIBLOCKADE ARTIFICIAL LATTICE HAMILTONIAN}
In this appendix, we will show the derivation of Eq.~(\ref{eq22}) in Ref.~\cite{wu2025perfect}. According to Fig.~\ref{fig5}(a), in the interaction picture, the Hamiltonian of the system is
\begin{equation}
\begin{aligned}
\hat{H}_{s} & =\hat{H}_{0}+\hat{H}_{d d}, \\
\hat{H}_{0} & =\left(\Omega_{1}|g\rangle_{1}\langle r| \mathrm{e}^{\mathrm{i} \Delta t}+\Omega_{2}|g\rangle_{2}\langle r| \mathrm{e}^{\mathrm{i} \Delta t}+\Omega_{3}|g\rangle_{3}\langle r| \mathrm{e}^{\mathrm{i} \Delta t}\right)+\text{H.c.}, \\
\hat{H}_{d d} & =V_{1}|r r g\rangle\langle r r g|+V_{1}|g r r\rangle\langle g r r|+V_{2}|r g r\rangle\langle r g r|+\left(2 V_{1}+V_{2}\right)|r r r\rangle\langle r r r|,
\end{aligned}    
\end{equation}
where $H_{0}$ is the Hamiltonian of the free atoms, $H_{dd}$ is the Hamiltonian of the interatomic interactions, and $V_1$ and $V_2$ represent the dipole-dipole interactions between nearest-neighbor and next-nearest-neighbor atoms, respectively.  First, by applying the unitary transformation \(e^{-iH_{dd} t}\)
 to Eq. (D1), we obtain the following three-atom Hamiltonian,
\begin{equation}
\begin{aligned}
H ={}& \Omega_1 \left( |ggg\rangle \langle rgg| e^{i\Delta t} + |grg\rangle \langle rrg| e^{i(\Delta - V_1)t} + |ggr\rangle \langle rgr| e^{i(\Delta - V_2)t} + |grr\rangle \langle rrr| e^{i(\Delta - V_1 - V_2)t} \right) \\
&+ \Omega_2 \left( |ggg\rangle \langle grg| e^{i\Delta t} + |rgg\rangle \langle rrg| e^{i(\Delta - V_1)t} + |ggr\rangle \langle grr| e^{i(\Delta - V_1)t} + |rgr\rangle \langle rrr| e^{i(\Delta - 2V_1)t} \right) \\
&+ \Omega_3 \left( |ggg\rangle \langle ggr| e^{i\Delta t} + |grg\rangle \langle grr| e^{i(\Delta - V_1)t} + |rgg\rangle \langle rgr| e^{i(\Delta - V_2)t} + |rrg\rangle \langle rrr| e^{i(\Delta - V_1 - V_2)t} \right) \\
&+ \text{h.c.},
\end{aligned}
\end{equation}
When the detunings, Rabi frequencies, and dipole–dipole interaction strengths satisfy
\begin{equation}
\Delta=V_1,
\qquad
V_{1,2}\gg\Omega_{1,2,3}.
\end{equation}
Under the rotating-wave approximation, two atoms can then be simultaneously excited to the Rydberg state, giving rise to the Rydberg anti-blockade effect. The entire three-atom system can consequently be described by the Hamiltonian of a five-site synthetic lattice:
\begin{equation}
\begin{aligned}
H={}&
J_1|rgg\rangle\langle rrg|
+J_2|rrg\rangle\langle grg|
+J_3|grg\rangle\langle grr| \\
&+J_1|grr\rangle\langle ggr|
+\mathrm{H.c.}
\end{aligned}
\end{equation}
Here, the Rydberg anti-blockade states $|rrg\rangle$ and $|grr\rangle$ represent two artificial lattice sites, and the coupling strengths between the lattice sites are given by
\begin{equation}
J_1=\Omega_2,
\qquad
J_2=\Omega_1,
\qquad
J_3=\Omega_3,
\end{equation}
To achieve perfect entanglement transfer, the parameters $J_i$ in the above equation must satisfy the following relation:
\begin{equation}\label{D6}
J_{i}=\frac{\Omega}{2}\sqrt{i(N-i)} .
\end{equation}
where \(N\) is the number of lattice sites and \(i\)is the atomic index.

\bibliography{ref}

\end{document}